\documentclass[11pt]{article}
\pdfoutput=1
 \usepackage{mciteplus}
 \usepackage{tikz}
 \usepackage{color}
 \usepackage{xcolor}
 \usepackage{comment}
 \definecolor{darkblue}{rgb}{0.1,0.1,.7}
 \usepackage[colorlinks, linkcolor=darkblue, citecolor=darkblue, urlcolor=darkblue, linktocpage,hyperfootnotes=false]{hyperref} 
\usepackage{epsfig}
\usepackage{graphicx}
\usepackage{cite}
\usepackage{amsfonts}
\usepackage{amssymb}
\usepackage{bm}
\usepackage{latexsym}
\usepackage{amsmath}
\numberwithin{equation}{section}
\def\bq{\begin{quote}}
\def\eq{\end{quote}}
 at 10truept

\newcommand{\calo}{{\cal O}}
\newcommand{\calh}{{\cal H}}

\newcommand{\beq}{\begin{equation}}
\newcommand{\eeq}{\end{equation}}
\newcommand{\beqa}{\begin{eqnarray}}
\newcommand{\eeqa}{\end{eqnarray}}
\newcommand{\bea}{\begin{eqnarray}}
\newcommand{\eea}{\end{eqnarray}}

\newcommand{\hf}{\frac{1}{2}}

\def\lesssim{~\mbox{\raisebox{-.6ex}{$\stackrel{<}{\sim}$}}~}
\def\roughly#1{\raise.3ex\hbox{$#1$\kern-.75em\lower1ex\hbox{$\sim$}}}

\begin{document}

\thispagestyle{empty}
\begin{titlepage}
  \bigskip

  \bigskip\bigskip

  \bigskip

\begin{center}
{\Large \bf {Quantum evolution of the Hawking state for black holes
}}
    \bigskip
\bigskip
\end{center}

  \begin{center}

 \rm {Steven B. Giddings\footnote{\texttt{giddings@ucsb.edu}} and Julie Perkins\footnote{\texttt{jnperkins@ucsb.edu}}
 }
  \bigskip \rm
\bigskip

{Department of Physics, University of California, Santa Barbara, CA 93106, USA}  \\
\rm

  \bigskip \rm
\bigskip
 
\rm

\bigskip
\bigskip

  \end{center}

\vspace{3cm}
  \begin{abstract}

We give a general description of the evolving quantum state of a Schwarzschild black hole, in the quantum field theory approximation.  Such a time-dependent description is based on introducing a choice of time slices.  We in particular consider  slices that smoothly cross the horizon, and introduction of ``stationary" such slices simplifies the analysis.  This analysis goes beyond standard derivations of Hawking radiation that focus on asymptotic excitations, and in particular gives an evolving state that is regular at the horizon, with no explicit transplanckian dependence, and that can in principle be generalized to incorporate interacting fields.  It is also expected to be useful in connecting to information-theoretic investigation of black hole evolution.
The description of the evolving state depends on the choice of slices as well as coordinates on the slices and mode bases; these choices give  different ``pictures" analogous to that of Schr\"odinger.
Evolution does have a simpler appearance in an energy eigenbasis, but such a basis is also singular at the horizon; evolution of regular modes has a more complicated appearance, whose properties may be inferred by comparing with the energy eigenbasis.  In a regular description, Hawking quanta are produced in a black hole atmosphere, at scales comparable to the horizon size.  This approach is also argued to extend to more general asymptotics, such as that of anti de Sitter space.  In the latter context, this analysis provides a description of the hamiltonian and evolution of a black hole that may be compared to the large-$N$ dynamics of the proposed dual CFT.

 \medskip
  \noindent
  \end{abstract}
\bigskip \bigskip \bigskip 

  \end{titlepage}

\section{Introduction}

Hawking radiation\cite{Hawk} is apparently one of the most mysterious phenomena in the world of physics.  While it appears to follow straightforwardly from the basic principles of local quantum field theory (LQFT) extended to curved spacetime backgrounds, its ultimate implications include an apparent internal contradiction among the basic principles of physics.  This ``black hole information problem," or perhaps more aptly, ``unitarity crisis," seems to point to the necessity of revising fundamental principles, in connection with understanding the foundations of quantum gravity.

While black holes (BHs) thus may play a key role in understanding these principles, the phenomenon of particle production in a nontrivial gravitational background is also important because of its role in the very early Universe, and in particular during a possible phase of inflationary expansion, which is believed to have created the fluctuations that lead to the large-scale structure in the visible matter distribution in the cosmos.  The seed fluctuations can be derived by methods closely parallel to those used to describe production of Hawking radiation.

In investigating the unitarity crisis, a lot of recent thought has focussed on the view that a more basic understanding of quantum information and its evolution in quantum gravity is important.  In particular, an important question is whether it is possible to think of a black hole and its environment as quantum subsystems, at least to a good approximation, of a larger quantum system, which evolve together in time.\footnote{For further discussion of this question, see \cite{SGsub}.}
This does appear to be the correct leading order picture, with possible small modifications that ultimately lead to a description consistent with unitarity.  

The original derivation\cite{Hawk} of Hawking radiation was based on an asymptotic description, analogous to that of the S-matrix:  it analyzed the asymptotic state of the radiation, but didn't directly describe the time evolution of the quantum state of the BH and its surroundings.  Thus, its connection to a description of an evolving quantum system is not direct.
Rederivations of the Hawking effect have largely followed in this vein, though various approaches have provided some additional information about the evolving quantum state.

A goal of this paper is to give a direct and more complete treatment of the evolving quantum state of a BH and its surroundings.  We will focus on this in the approximation where it is described within LQFT, and thus will not attempt to describe the more complete dynamics that is ultimately expected to be unitary.  However, this approximate evolution plays an important background role in describing this unitary dynamics, if the latter is a correction to this evolution that is in certain regimes a small correction.  Another role for the present description is in treating Hawking radiation for interacting theories; treatment of interactions in Hawking's original derivation is problematic due to its reliance on evolution via free mode propagation.

Specifically, we will describe the time-dependent evolution in a picture analogous to standard Schr\"odinger picture, by making a choice of time slices that is regular across the horizon, and deriving the resulting evolution of the quantum wavefunction.  This was previously done for two-dimensional black holes in \cite{SEHS,SE2d}; earlier work on dynamical evolution on such slices includes \cite{MeWe1,MeWe2,BHS,HoSi}.  
  Description of this evolution also relies on choosing coordinates on the time slices, and a basis of modes.

In outline, we begin in the next section by paremeterizing such slicings, and describing the corresponding Arnowitt-Deser-Misner (ADM)  parameterization of the metric.  The following section then derives the hamiltonian for scalar matter in such a slicing, and the canonical quantization of the theory.  A choice of modes  leads to a Fock construction of the Hilbert space, and construction of the evolution operator acting on it.   In fact, there are many such descriptions of the evolving quantum state, which depend on the specific choice of slices, coordinates, and mode basis; these are analogous to different ``pictures" of the evolution (and are expected to be equivalent).  

In a free theory, this evolution can be simplified by in particular using energy eigenstates for the mode basis.  Section four describes such modes (and Appendix A finds an explicit form of them in $D=4$ spacetime dimensions in terms of Heun functions) and outlines their important properties.  While the evolution is simplified in such a basis, the basis is {\it singular}.  This connects to Hawking's and related derivations, where modes are traced back to ultraplanckian wavelengths near the horizon.  The resulting transplanckian behavior has served as a source of concern and confusion in the literature, but from this viewpoint just arises from choice of  a singular basis to describe the state.

These issues may be avoided, as in the next section, by  instead working with a regular basis.  This does lead to a more complicated description, but one that in principle exhibits evolution without any explicit reference to transplanckian excitations.  This section in particular finds the form of the resulting hamiltonian.  

Section six then puts the previous treatments together to give a description of the evolution of the quantum state, which we call the Hawking state, resulting from a collapsing BH. Again, this can be done in terms of a regular mode basis, but at the price of a more complicated evolution law.  We can, however, learn about its structure by comparing it to the singular and simpler description in terms of energy eigenmodes.  In particular, one can see the familiar behavior of asymptotic Hawking excitations, as well as of the internal partner excitations and pairing between inside and outside excitations.  We also discuss the internal evolution on ``nice slices," and exhibit a ``frozen" description of the internal state, in a picture that results from particular choices of internal coordinates.  

The final section outlines the extension of these results to interacting theories, and to situations with different asymptotic metrics besides that of Minkowski.  It in particular discusses the case of anti de Sitter space.  Here, the same quantization procedure is argued to yield a hamiltonian and evolution that furnishes, in AdS/CFT language, a leading order large-$N$ description of quantum evolution of a BH, as well as $1/N$ corrections arising from interactions.  Here, again, this is not expected to yield evolution that is ultimately unitary, connecting to the question of the form of additional corrections needed to unitarize the dynamics.

\section{Geometry and time slicings}\label{Slicing}

We begin by describing the geometry of a BH, and time slices of that geometry.

\subsection{Schwarzschild parameterizations} 
The standard form of the Schwarzschild metric is
\beq
\label{Schmet}
ds^2 = -f(r) dt^2 + \frac{dr^2}{f(r)} + r^2 d\Omega^2
\eeq
where for $D>3$ dimensions
\beq
f(r) = 1-\left(\frac{R}{r}\right)^{D-3}\ ;
\eeq
here $R$ is the horizon radius.  Both for describing field propagation and for giving a smoother description of the geometry, it
 is useful to introduce conformal coordinates for the $t,r$ plane, by defining $r_*(r) = \int dr/{f(r)}$ so that
\beq
\label{rstarmet}
ds^2 = f(r)(- dt^2 +dr_*^2) + r^2 d\Omega^2.
\eeq
For example in $D=4$, 
\beq
r_* = \int \frac{dr}{1-R/r} = r-R+R\ln\left(\frac{r}{R}-1\right)\ ,
\eeq
up to an overall additive constant.  
Then, for the exterior of the horizon, we can define left/right moving Eddington--Finkelstein coordinates 
\beq\label{xpdef}
x^\pm = t\pm r_*\ .
\eeq

While the Schwarzschild coordinates \eqref{Schmet} and the definition of $r_*$ are clearly singular at the horizon, the latter definition can be extended to $r<R$;  for example in $D=4$
\beq
r_* =   r-R+R\ln\left(1-\frac{r}{R}\right) \pm\pi i R\ .
\eeq
The sign reversal of $f$ in \eqref{rstarmet} indicates that $r_*$ plays the role of a time coordinate for $r<R$, and left/right moving coordinates for the interior are
\beq
{\hat x}^\pm = r_*\pm t\ .
\eeq

The metric \eqref{Schmet} is of course smooth across the horizon, and this can for example be exhibited by working in the incoming Eddington--Finkelstein coordinates, $(x^+,r)$.  This gives
\beq
\label{EFmet}
ds^2 = -f(r) dx^{+2} + 2dx^+ dr + r^2 d\Omega^2
\eeq
which smoothly covers the region $r>0$.  The time translation invariance is also inherited by this form of the metric, and becomes invariance under
\beq\label{ttrans}
x^+\rightarrow x^+ + {\rm constant}\ .
\eeq
This invariance plays an important role in the dynamics.

\subsection{Slicings and ADM description}
\label{Slicesec}

In order to describe dynamical evolution, we can provide a foliation of the geometry \eqref{Schmet}, \eqref{EFmet} by time slices.  Such a foliation can in general be parameterized as 
\beq\label{sliceparam}
x^\mu= {\cal X}^\mu(t,x^i)
\eeq
where now $t$ labels slices of the foliation, and $x^i$ is a spatial coordinate. In a Schwarzschild background, one expects the description to be simplest for a foliation respecting the spherical symmetry, so that
\beq\label{xrparam}
x^+={\cal X}^+(t,x)\quad,\quad r={\cal X}^r(t,x)\ ,
\eeq
independent of angles, with general radial coordinate $x$, and using the standard angular coordinates.  We also can anticipate some simplifications for slicings that  respect the translation symmetry \eqref{ttrans}, and these take the general form
\beq\label{statgen}
x^+= t+ s(x)\quad,\quad r=r(x)\ ,
\eeq
which we refer to as a ``stationary slicing" \cite{NVU,SEHS}.
Their specification particularly simplifies if we use $r$ as the radial coordinate,
\beq\label{statS}
x^+= t+S(r)\ .
\eeq

\begin{figure}[!hbtp] \begin{center}
\includegraphics[width=10cm]{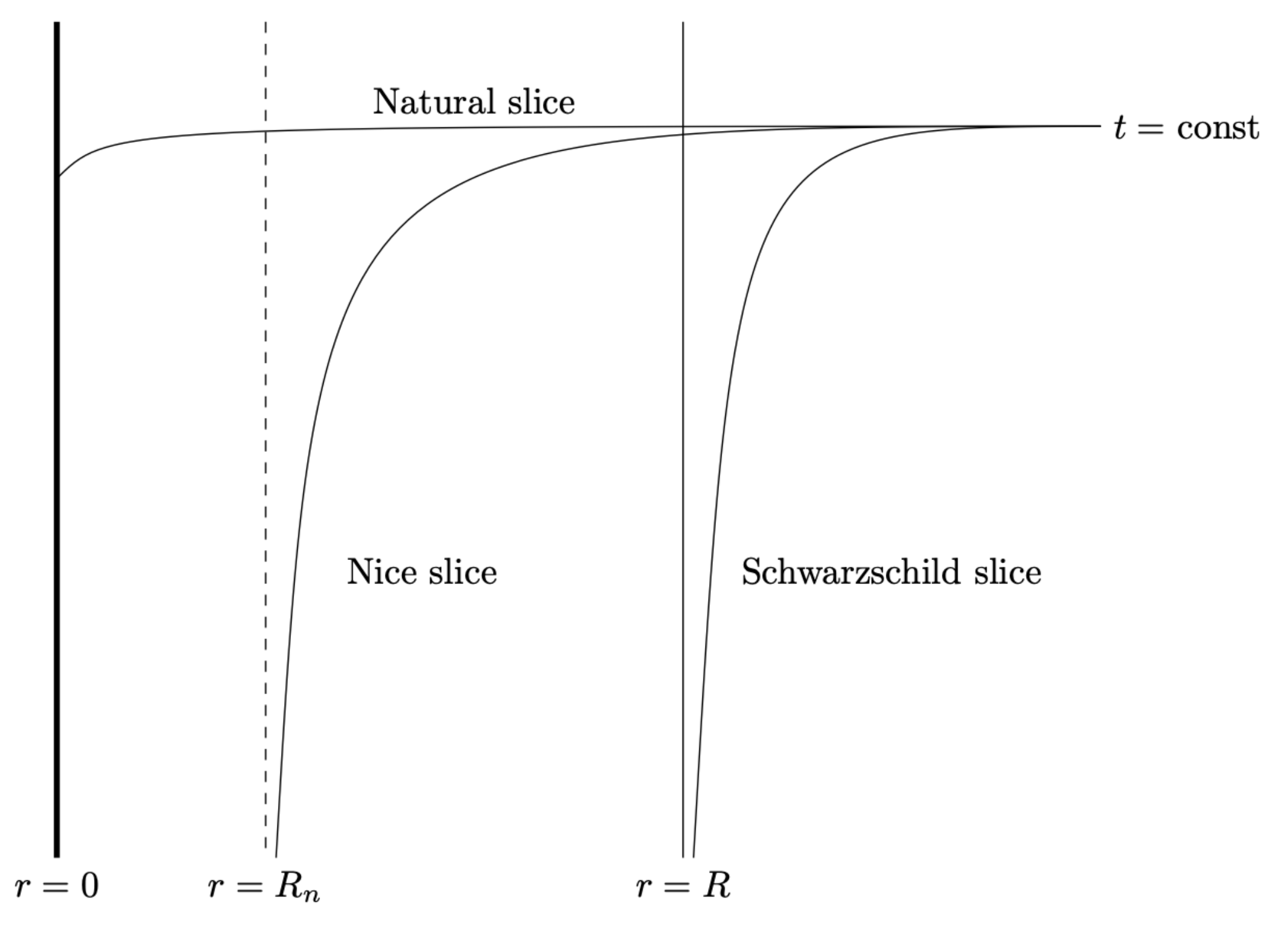}
\end{center}
\caption{Shown in an Eddington-Finkelstein diagram are different kinds of slices. In addition to the familiar Schwarzschild slices, there are nice slices, asymptoting to a constant $r=R_n$, and natural slices which reach $r=0$.  These slices all asymptote to constant Schwarzschild time slices at $r=\infty$.  The full family of slices of the geometry is found by translating one of these slices vertically in the figure, corresponding to a time translation in Schwarzschild time.}
\label{Figslices}
\end{figure}

This family of slicings unifies various descriptions of the Schwarzschild spacetime, and the form of the ``slice function" $S(r)$ plays a key role.  For example, from the coordinate definition \eqref{xpdef}, we see that $S(r)=r_*(r)$ gives the Schwarzschild time slicing, relevant for observers who stay outside the horizon.  Slicings that cross the horizon -- for example describing observations of a family of observers, some of whom enter the BH -- arise from slice functions that are smooth at the horizon.  Freely falling such observers will reach $r=0$, and that is naturally described by a family of ``natural slices" for which $S(r)$ is finite there.  A very simple example is $S(r)=r$, corresponding to ``straight slices,"  which lead to some simplifications.  However, such natural slices do not give good Cauchy slices, since they cease to describe excitations that have reached $r=0$, in the absence of a supplementary description there.  Good Cauchy slices can, however, be specified by using an $S(r)$ that asymptotes to minus infinity at some finite $r=R_n<R$.  This gives an example of the construction of ``nice slices," such as were described by \cite{Waldnice,LPSTU}.  These different kinds of slices are shown in Fig.~\ref{Figslices}.

Evolution over a general  slicing is conveniently described by using ADM variables\cite{ADM} for the metric,
\beq
\label{ADMmet}
ds^2 = -N^2 dt^2 +q_{ij}(dx^i + N^i dt)(dx^j + N^j dt)\ .
\eeq
The lapse $N$, shift $N^i$, and spatial metric $q_{ij}$ are dependent on the spatial coordinates, and we also define
$N_i =q_{ij} N^j$. For the Schwarzschild metric in a stationary slicing described by \eqref{statS}, these functions become\cite{NVU}
\beq\label{ADMvar}
N^2 = \frac{1}{S'(2-f S')},\quad \quad N_r = 1-f S',\quad\quad q_{rr} = S'(2-f S')\ ,
\eeq
with $S'=dS/dr$, and
with the remaining components $q_{ij}$ of standard angular form.  This is also readily generalized for a more general stationary radial coordinate, 
\eqref{statgen}.
It is also useful to have the unit normal to the time slices, which takes the general form 
\beq\label{normdef}
n^{\mu} = (1,-N^i)/N .
\eeq

\section{Schr\"odinger Description}
\label{Schsec}

\subsection{Canonical quantization}

Our goal is to describe the evolution of the quantum state of a BH.  
Of course, emitted Hawking radiation changes the mass of the BH, making the geometry non-stationary.  However, the average time to emit a Hawking quantum of energy $\sim 1/R$ is $R$.  This means that the fractional change in the mass over the characteristic emission time is $R dM/Mdt\sim 1/RM \sim 1/S_{BH}$, where $S_{BH}$ is the Bekenstein-Hawking entropy.  This small parameter justifies using the stationary approximation over times $\ll R S_{BH}$ for large BHs.  We will focus on this approximation and describe evolution of quantum fields on the stationary BH background, leaving treatment of quantum backreaction for future work.
For simplicity, we consider  evolution of a free massless scalar field $\phi$ in a $D\geq 4$ dimensional Schwarzschild background. Working in ADM variables, with a general slicing, the action takes the form 
\beq\label{Sact}
S=-\hf \int d^D x \sqrt{|g|} (\nabla\phi)^2 =  \frac{1}{2} \int dt d^{D-1}x \sqrt q N \left[ (\partial_n\phi)^2 - q^{ij}\partial_i \phi \partial_j \phi\right]\ ,
\eeq
where we have defined the normal derivative $\partial_n\phi=n^\mu\partial_\mu \phi$.  
The canonical momentum is then defined by
\beq\label{canonmom}
\pi= \frac{1}{\sqrt q}\frac{\delta S}{ \delta \dot\phi} = \frac{1}{N}\left(\partial_t \phi - N^i\partial_i\phi\right) = \partial_n\phi\ .
\eeq
Using this, we can write the canonical form of the action
\beq
S=\int dt d^{D-1}x \sqrt q \left(\pi \dot\phi  - \calh\right)\ ,
\eeq
where the hamiltonian is
\beq\label{ADMH}
H =  \int d^{D-1} x \sqrt q \calh= \int d^{D-1} x \sqrt q \left[ \frac{N}{2}\left(\pi^2+q^{ij}\partial_i\phi \partial_j\phi\right) + \pi N^i \partial_i\phi \right] \ .
\eeq 
Quantization proceeds via the equal time canonical commutation relations
\beq\label{CCRs}
[\pi(x^i, t), \phi(x^{i\prime}, t)]=-i \frac{\delta^{D-1}(x-x')}{\sqrt{q}}\ .
\eeq

Note that the hamiltonian \eqref{ADMH} depends on the choices of both foliation and spatial coordinate in the general expression
\eqref{sliceparam}, through the dependence of $N$ and $N^i$ on these choices.  This results in what can effectively be described as different ``pictures" for the evolution, generalizing the choice of Heisenberg or Schr\"odinger picture, as also discussed in \cite{SE2d}.  In a given such Schr\"odinger picture, we take the field and momentum operators to be time independent, and all evolution to be in the state.\footnote{In a time-dependent background, there are further subtleties with Schr\"odinger picture discussed in \cite{ToVa,CMOV,CFMM,AgAs,MuOe}. The present work avoids these with the time-independent background and slicing.}  

Description of the evolution also depends the choice of a mode basis, given by specifying a complete set of pairs of functions $\gamma_I(x^i)=(\phi_I(x^i),\pi_I(x^i))$.  Such a pair gives Cauchy data for a solution $\phi_I(x^i,t)$.  For quantization, one also specifies  a choice of complex structure 
\cite{More,Kay,AgAs} that separates these into ``positive frequency" modes $\gamma_A(x^i)=(\phi_A(x^i),\pi_A(x^i))$ and conjugate ``negative frequency" modes $\gamma_A^*(x^i)$.  
The inner product of two such sets of Cauchy data is 
\beq
(\gamma_1, \gamma_2)= i\int d^{D-1}x \sqrt{q}(\phi_1^*\pi_2-\pi_1^*\phi_2)\ ,
\eeq
and extends to an inner product between solutions,  
\beq\label{InnerProduct}
(\phi_1,\phi_2) = i \int d^{D-1} x \sqrt{q} n^\mu \phi_1^*\overleftrightarrow{\partial_\mu} \phi_2\ ,
\eeq
which is conserved by the equations of motion.  Different such choices of modes also lead to different pictures.

The Schr\"odinger picture field operators, in a given such picture, can be expanded as
\beq\label{opexps}
\phi(x^i) = \sum_A \big[a_A \phi_A(x^i)+a_A^{\dagger} \phi_A^*(x^i)\big]\quad ,\quad \pi(x^i) = \sum_A \big[a_A\pi_A(x^i)+a_A^{\dagger} \pi_A^*(x^i)\big]\ .
\eeq
If the mode basis is orthonormal,
\beq
(\gamma_A,\gamma_B) = \delta_{AB}\quad , \quad (\gamma_A,\gamma_B^*) = 0\ ,
\eeq 
the canonical commutators imply the commutation relations
\beq
[a_A, a_B^{\dagger}] = \delta_{AB}\quad ,\quad  [a_A, a_B]=[a_A^{\dagger}, a_B^{\dagger}] = 0\ .
\eeq
A Fock space basis for the Hilbert space then arises by acting with the creation operators $a_A^\dagger$ on the vacuum $|0\rangle$ annihilated by the $a_A$.  

Schr\"odinger picture evolution is then described by the action of the time evolution operator,
\beq
U(t_2, t_1)= \exp{\bigg[-i\int_{t_1}^{t_2} H dt\bigg]}\ ,
\eeq
determined by the hamiltonian \eqref{ADMH}, on the state.  And, for example, if initially the state is the vacuum state $|0\rangle$ (in a particular basis), it will not necessarily remain in that state, since the hamiltonian in general creates additional excitations.

\subsection{Hamiltonian and pictures}\label{HandP}

The hamiltonian \eqref{ADMH} can be written in other forms which are useful in describing evolution. We begin by using  the explicit definition of the 
momentum to obtain from \eqref{ADMH}
\beq\label{LapseForm}
H= \int d^{D-1}x \sqrt{q} \left[\frac{1}{2N} (\partial_t\phi)^2+\frac{N}{2}g^{ij} \partial_i \phi \partial_j \phi \right].
\eeq 
We can rewrite this in terms of a vector $\xi=\partial_t$, which connects points on neighboring slices with equal spatial coordinates.
In component form, this becomes from \eqref{sliceparam}
\beq
\xi^{\mu} = \frac{\partial {\cal X}^{\mu}}{\partial t}\Big\vert_{x^i},
\eeq 
at fixed spatial coordinate $x^i$. Then, using the stress tensor for the minimally coupled scalar field, the  Hamiltonian becomes
\beq\label{StressTensorForm}
H_{\xi}= \int d^{D-1}x \sqrt{q} n^{\mu} \xi^{\nu} T_{\mu\nu}\ ,
\eeq
where $n^{\mu}$ is the unit normal \eqref{normdef}. 

This form of the hamiltonian also exhibits the dependence both on the slicing and on the choice of spatial coordinate $x^i$ along the slices; for example a redefinition $x^{i\prime}(x^j,t)$ changes $\xi^\mu$ and thus $H_\xi$.  The different choices of $\xi$ define different Hamiltonians and Schr\"odinger pictures,  which lead to distinct descriptions of the evolution.  In particular, the Hamiltonian $H_{\xi}$ is conserved when $\xi$ is a Killing vector, since the stress tensor is also conserved. Such a Killing vector is present after the  matter has collapsed to form a BH.  The expression \eqref{StressTensorForm} can alternately be derived in the covariant canonical formalism; see {\it e.g.} appendix B of \cite{DoGi2} for a review.

Another  useful expression can be found by putting the hamiltonian \eqref{LapseForm} in a form which resembles the conserved inner product \eqref{InnerProduct}. We will connect to the inner product by introducing the canonical momentum $\pi=\partial_n\phi$ into \eqref{LapseForm}. Replacing one of the time derivatives using \eqref{canonmom}, the Hamiltonian becomes
\beq
H= \int d^{D-1}x \sqrt{q}
\left(\frac{1}{2}\partial_t \phi \partial_n\phi + \frac{1}{2}\partial_t\phi\frac{N^i}{N}\partial_i  \phi  +\frac{N}{2}g^{ij} \partial_i \phi \partial_j \phi \right).
\eeq 
Integrating the last two terms by parts with respect to  $x^i$, and neglecting the boundary term, the expression becomes
\beq
H= \int d^{D-1}x \sqrt{q} \left[\frac{1}{2}\partial_t \phi \partial_n\phi -\frac{\phi}{2\sqrt{q}} \partial_i \left(\sqrt{q} N g^{i \mu} \partial_{\mu} \phi \right)\right].
\eeq 
Finally, using the equation of motion to rewrite the second term, the Hamiltonian takes the simplified form
\beq\label{IPform}
H= \frac{1}{2} \int d^{D-1}x \sqrt{q} \left(\partial_t \phi \partial_n\phi -\phi \partial_t \partial_n \phi -\partial_n\phi \phi \frac{\partial_t q}{2q} \right).
\eeq 
Comparing the above equation to \eqref{InnerProduct}, we see that for time-independent metric coefficients it reduces to the inner product $i(\phi^*,\xi^\mu\partial_\mu \phi)/2$.  This will be useful in the following analysis, particularly when considering special choices of modes.

\section{Energy eigenmodes and their evolution}\label{EnergyEmodes}

The time-translation symmetry \eqref{ttrans} suggests expanding in a basis of modes that correspond to eigenstates of the time translation generator, which is  $\partial_{x^+}$, or $\partial_t$ in the slice coordinates. We begin by separating off the angular coordinates using\footnote{For spacetime dimension $D>4$, the spherical harmonics involve multiple angular quantum numbers $m_i$; for notational simplicity, we use the four-dimensional notation $Y_{lm}$ in the following discussion.}
\beq\label{phisep}
\phi(x)\sim u_{l}(x^+,r) \frac{Y_{lm}(\Omega)}{r^{D/2-1}}\ .
\eeq
Then $u_{l}(x^+,r)$ obeys the equation
\beq\label{rxpeq}
\partial_r\left(2\partial_+ u + f\partial_r u\right) - V_l(r)u =0\ 
\eeq
with potential
\beq
V_l(r)= \left(\frac{D}{2}-1\right)^2 \frac{R^{D-3}}{r^{D-1}} +\frac{l(l+D-3)+(D-2)(D-4)/4}{r^{2}}\ ;
\eeq
{\it e.g.} in the case $D=4$, 
\beq\label{4dpot}
V_l(r) = \frac{R}{r^3} + \frac{l(l+1)}{r^2}\ .
\eeq
Eigenfunctions of the time translation symmetry then take the form
\beq\label{Eefcns}
e^{-i\omega x^+}u_{\omega l }(r) \frac{Y_{lm}}{r^{D/2-1}}\ ,
\eeq
or in the $t,r$ coordinates arising from a stationary slicing \eqref{statS}
\beq\label{Trsolns}
e^{-i\omega t}U_{\omega l}(r) \frac{Y_{lm}}{r^{D/2-1}}\quad ,\quad U_{\omega l}(r)= e^{-i\omega S(r)}u_{\omega l}(r)\ .
\eeq

Again, 
the special case of a Schwarzschild slicing corresponds to $S(r)=r_*(r)$, and  the equation \eqref{rxpeq} can be simplified by defining
\beq\label{rescm}
 g_{\omega l}(r)=e^{-i\omega r_*} u_{\omega l}(r)
\eeq
and becomes
\beq\label{diffeq}
\frac{d^2 g_{\omega l} }{d r_*^2}  + \left[\omega^2-f(r)V_l(r)\right] g_{\omega l} =0\ .
\eeq
At large $r$, or near the horizon, the effective potential $fV_l$ vanishes, and we have 
\beq\label{gasymp}
g_{\omega l}\sim \exp\{\pm i \omega r_*\}\ .
\eeq
Solving for $g_{\omega l}$ becomes a well-known barrier penetration problem.  Inside the horizon, we also find the behavior \eqref{gasymp} near the horizon, $r_*\rightarrow-\infty$.  The general internal solutions are difficult to find, but one indicator of their behavior is their WKB approximation, which has the form
\beq
g_{\omega l} \sim e^{\pm i \int dr_* \sqrt{\omega^2 -fV_l}}\ .
\eeq
This approximation in particular fails as $r$ approaches zero ($r_*\rightarrow-R$), but does illustrate the rapidly varying nature of the solutions.

The behavior of energy eigenmodes can be further understood by examining the differential equation \eqref{rxpeq}. The equation has regular singular points at $0$ and $R e^{2 \pi i n/ (D-3)}$ for integers $n = 0, 1,..., D-4$, and an irregular singular point at infinity (see Appendix~\ref{Happend}). The  solutions are not known in general, however, in $D=4$ \eqref{rxpeq} can be transformed into the confluent Heun equation. In this case of  $D=4$ Schwarzschild BHs the solutions to the Heun equation are well known and have been widely studied in the literature, and incoming and outgoing modes are classified by their behavior near the singular points. We will not need the detailed behavior of the solutions to  \eqref{rxpeq} in order to see that they define a basis, but their asymptotic behaviors will be important.  

One can describe quantization using a basis of such solutions. One way to characterize solutions is in terms of their behavior at $t\rightarrow-\infty$.  There are incoming solutions from $r=\infty$, but also can be (singular) solutions that asymptote to $r=R$ (or $r_*=-\infty$) in the far past.  Specifically, we can introduce the following basis:
\begin{itemize}
\item{$\tilde u_{\omega l}$}: these ``in" modes are modes such that the coefficient of $e^{i\omega r_*}$ vanishes near the horizon, {\it i.e.} $\tilde g_{\omega l}\sim e^{-i\omega r_*}$.  Eq.~\eqref{rescm} then shows that these modes are non-singular at the horizon; they have purely ingoing behavior there, $\phi\sim e^{-i\omega x^+}$.  At $r_*=r=\infty$ they have both an ingoing piece, which may be normalized to unity, and a reflected outgoing piece.  The internal part of the solution also is purely ingoing at the horizon, but takes a more general form for finite $r_*$.    
\item{$u_{\omega l}$}: these ``up" modes are modes that in $g_{\omega l}$ have nonvanishing coefficient of $e^{i\omega r_*}$, taken to be unity, as $r\rightarrow R_+$, but vanishing coefficient as $r\rightarrow R_-$, and with purely outgoing wave $e^{i\omega r_*}$ at $r_*\rightarrow \infty$.  These both give behavior $\phi\sim e^{-i \omega x^-}$.  These also have a reflected $e^{-i\omega r_*}$ piece at the horizon, which continues to the interior similarly to the previous case. 
\item{$\hat u^*_{\omega l}$}:  these ``inside" modes are modes that in $\hat g_{\omega l}$ have nonvanishing coefficient of $e^{i\omega r_*}$, taken to be unity, as $r\rightarrow R_-$, and which vanish outside the horizon.  
Thus near the horizon $\phi\sim e^{i\omega \hat x^-}$; 
for finite internal $r_*$, they  have general behavior.
\end{itemize} 
In the exterior region $r>R$,  the ``in" and ``up" modes  correspond to those of \cite{ChMi}, also discussed in \cite{GiSh2}, but we have also described the interior continuations of these solutions, which have an  ingoing part.  We have also introduced the ``inside" modes.  
Of course, the ``up" and ``inside" modes are singular at the horizon, due to the singularity at $r=R$ in the definition of $r_*$.   The corresponding full solutions are denoted
\beq\label{EEigenmodes}
\phi_{\omega l m}(x^i,t) 
= e^{-i\omega x^+} u_{\omega l}(r)  \frac{Y_{lm}}{r^{D/2-1}}\ ,
\eeq
and likewise for $\tilde\phi_{\omega lm}$, $\hat\phi^*_{\omega lm}$. 

These solutions, written in terms of $g_{\omega l}$ via  
 \eqref{rescm}, are of course orthogonal under the conserved inner product \eqref{InnerProduct}
unless $l,m$ match, in which case the product becomes
\beq\label{EMIP}
(\phi_1,\phi_2)= \int \frac {dr}{f} \left[(\omega_1+\omega_2) g_{\omega_1l}^* g_{\omega_2l} - i f(1-fS') g_{\omega_1l}^* \overleftrightarrow{\partial_r} g_{\omega_2l} \right] e^{i(\omega_1-\omega_2)[t+S(r)-r_*]}\ .
\eeq
One can see that the sets of modes $\phi_{\omega lm}$, $\hat \phi_{\omega lm}$, $\tilde \phi_{\omega lm}$ are mutually orthogonal by considering a general localized wavepacket of such modes.  In the far past, $t\rightarrow-\infty$, this wavepacket will localize near the horizon for $\phi_{\omega lm}$, $\hat \phi_{\omega lm}$, and near infinity for $\tilde \phi_{\omega lm}$, and so the inner product vanishes.  The modes in a given set 
are orthogonal for $\omega_1\neq \omega_2$, since otherwise \eqref{EMIP}  would contradict the time-independence of $(\phi_1,\phi_2)$.  
One also finds by examining their near-horizon behavior that the modes $\hat \phi^*_{\omega lm}$ are {\it negative} norm, and so their conjugates $\hat \phi_{\omega lm}$ are  positive norm solutions.\footnote{We also take $m\rightarrow-m$ in our definition, so that $\hat \phi_{\omega l m} \propto Y_{lm}$.}  Our normalization convention is $(\phi_{\omega l m},\phi_{\omega' l' m'}) = 4\pi \omega \delta(\omega-\omega') \delta_{ll'}\delta_{mm'}$, and similarly for $\hat \phi$, $\tilde \phi$.

The expansion of the field in terms of the modes  inside and outside of the horizon takes the form
\beq
\phi(x^i,t)= \sum_{lm}\int_0^\infty \frac{d\omega}{4\pi\omega}\left(b_{\omega l m} \phi_{\omega l m} + \tilde b_{\omega lm} \tilde\phi_{\omega l m} + \hat b_{\omega lm} \hat  \phi_{\omega l m} + h.c.\right)
\eeq
and may be, for example, evaluated at $t=0$, in a given slicing, to give the Schr\"odinger picture operator \eqref{opexps}.
This expansion also may be expressed compactly as
\beq\label{phiexp}
\phi= \sum_{A} b_A \phi_A + h.c.\ 
\eeq
where the integral over frequencies has been included in the general sum over modes labeled by $A$.
Note that for the purely internal $\hat \phi_{\omega lm}(r)$ modes, the frequencies have the opposite sign, in accord with the above definitions.   

As expected, the hamiltonian greatly simplifies in this basis.  From \eqref{IPform} we found for a stationary slicing 
\beq\label{Hinnerp}
H=\frac{1}{2}\int d^{D-1}x \sqrt{q} n^{\mu} [\partial_t\phi \partial_{\mu} \phi - \phi \partial_t \partial_{\mu} \phi]= \frac{1}{2} (\phi^*, i\partial_t \phi) \ .
\eeq
Then using 
\beq\label{tderiv}
\partial_t\phi(x)= -i\sum_{A} \omega_A b_A \phi_A + h.c.\
\eeq
and the orthogonality between modes, we find
\beq
H= \sum_{A} \omega_A b_A^{\dagger}b_A (\phi_A,\phi_A) +H_0 \,
\eeq 
where $H_0$ is the normal ordering constant
\beq
H_0 = \sum_A \frac{\omega_A}{2}  [b_A, b_A^{\dagger}]  (\phi_A,\phi_A) \ .
\eeq 
Returning to a more explicit labeling of the mode sum, the Hamiltonian becomes
\beq\label{EMham}
H= \sum_{lm} \int \frac{d\omega}{4 \pi \omega} \omega (b_{\omega lm}^{\dagger}b_{\omega lm} - \hat b_{\omega lm}^{\dagger} \hat b_{\omega lm} + \tilde b_{\omega lm}^{\dagger}\tilde b_{\omega lm}  ) +H_0 \ .
\eeq 
This has the same form as the $D=2$ case discussed  in \cite{SEHS,SE2d}, including both chiralities of the modes, as a result of the spherical symmetry of the spacetime.  One clearly sees that the ``inside" modes have negative energies for this hamiltonian.  

Of course the simplicity of the hamiltonian \eqref{EMham} is somewhat illusory, since the specification of a good initial state, {\it e.g.} with a regularity condition at the horizon, is rather more complicated in this basis, as was clearly illustrated in the 2d case in \cite{SEHS,SE2d}.  An alternate way to describe such a regular state is to work directly in terms of modes that are regular at the horizon, to which we now turn.

\section{Regular modes and their evolution}

To give a treatment of evolution respecting regularity at the horizon, it is most natural to consider a mode basis that is regular there.  
If we consider a general stationary slicing \eqref{statgen}, a mode basis may be specified by giving pairs of functions
$(\phi_A(x^i),\pi_A(x^i))$, with $A$ a basis label, on those slices.  These also provide Cauchy data for a corresponding solution that evolves forward from a given slice.  

\subsection{Properties of modes}

We have found that the ``in" energy eigenmodes are regular at the horizon, and so can be used to provide a regular basis, but the ``up" and ``inside" modes are singular there.  However, as two-dimensional examples illustrate\cite{SEHS,SE2d}, we expect to be able to  also find a mode basis that is regular at the horizon by combining these latter two.

Specifically, working with initial data on a slice which may be chosen to be at $t=0$, the space $\mathcal{H}_{in}= \text{Span}\{(\tilde{\phi}_{\omega lm}(x^i,0),\tilde{\pi}_{\omega lm}(x^i,0))\}$ describes ``in" modes.  This is orthogonal to the spaces 
$\mathcal{H}_{up}= \text{Span}\{({\phi}_{\omega lm}(x^i,0),{\pi}_{\omega lm}(x^i,0))\}$
and $\mathcal{H}_{inside}= \text{Span}\{(\hat{\phi}_{\omega lm}(x^i,0),\hat{\pi}_{\omega lm}(x^i,0))\}$, corresponding to the ``up" and ``inside" modes, where here the $\pi$'s are derived from the corresponding solutions described in the previous section using \eqref{canonmom}.  The orthogonality of the Cauchy data extends to orthogonality of the solutions.  We  expect to be able to combine elements of $\mathcal{H}_{up}$ and $\mathcal{H}_{inside}$ to give regular modes at the horizon.  

Explicitly, we  expect to be able to find regular modes which are determined by Cauchy data
\bea\label{Bogreln}
 \phi_{klm }(x^i,0)&= \int d\omega (\beta^+_{k\omega l} \phi_{\omega lm}+\beta^-_{k\omega l} (-1)^m\phi^*_{\omega l,-m}+
   \hat{\beta}^+_{k\omega l} \hat{\phi}_{\omega lm}+
   \hat{\beta}^-_{k\omega l} (-1)^m \hat{\phi}^*_{\omega l,-m})\ ,\cr
 \pi_{klm }(x^i,0)&= \int d\omega (\beta^+_{k\omega l} \pi_{\omega lm}+\beta^-_{k\omega l} (-1)^m\pi^*_{\omega l,-m}+
   \hat{\beta}^+_{k\omega l}  \hat{\pi}_{\omega lm}+
   \hat{\beta}^-_{k\omega l} (-1)^m \hat{\pi}^*_{\omega l,-m})\ ,
\eea
where $k$ is a continuous quantum number, and where it is understood that the up and inside mode functions contain factors of $\theta(r-R)$ and $\theta(R-r)$, respectively.  The regularity condition at the horizon enforces conditions on the Bogolubov coefficients $\beta^+, \beta^-,\hat \beta^+,\hat \beta^-$.  

We can then think of the modes $(\phi_{klm},\pi_{klm})$ as spanning a space  $\mathcal{H}^R_{up}\subset {\mathcal{H}}_{up} \oplus \mathcal{H}_{inside}$, which inherits its orthogonality to $\mathcal{H}_{in}$ from ${\mathcal{H}}_{up}$ and $ \mathcal{H}_{inside}$.  We expect the corresponding set $u_{kl}(r)$ and their complex conjugates to form a complete basis of functions of $r$, and
the functions \eqref{Bogreln} and their conjugates to form a basis for Cauchy data of regular solutions.
Of course, the corresponding solutions $\phi_{klm}(x^i,t)$ will then have non-trivial time dependence, as in the $D=2$ case in \cite{SEHS}. In $D>2$ dimensions the equation of motion also  includes an effective potential, which leads to mixing between right and left moving modes and more complicated solutions.

It appears difficult to give explicit expressions for such regular up bases, and that they are most easily treated approximately.  They can also be thought of as being specified by different characteristics.  One is that a localized wavepacket superposition of such solutions is expected to increasingly localize in the vicinity of $r=R$ in the far past, becoming singular in the infinite past, as is  seen in the simpler 2d example\cite{SE2d}.  
We may alternately imagine defining the modes by specifying regular functions $\phi_{klm}(x^i)$, and then choosing corresponding $\pi_{klm}(x^i)$ so that the modes are orthogonal to those in $\calh_{in}$ and satisfy an appropriate condition corresponding to a choice of ``positive frequency" (in the nomenclature of Sec.~\ref{HandP}), but this appears not to give conditions that are simple to solve.  Finally, these modes can be specified by requiring that they be regular at the horizon and that their evolution $u_{kl}(x^+,r)$ (compare \eqref{phisep}) be purely outgoing at $r=\infty$ for all $x^+$ or time.

Using a general such regular basis, the field and momentum operators can be expanded as
\bea\label{Fexp}
\phi(x^i)&=&\sum_{lm} \int_0^{\infty}\frac{dk}{4\pi k} \left(a_{klm}\phi_{klm}+\tilde{a}_{klm}\tilde{\phi}_{klm}+h.c.\right)\ ,\cr
\pi(x^i)&=&\sum_{lm} \int_0^{\infty}\frac{dk}{4\pi k} \left(a_{klm}\pi_{klm}+\tilde{a}_{klm}\tilde{\pi}_{klm}+h.c.\right)
\eea 
where both $\phi_{klm}$ and $\tilde{\phi}_{klm}$ are regular at the horizon. 

\subsection{Evolution of regular modes}

Using \eqref{Hinnerp} and \eqref{Fexp}, the Hamiltonian for regular modes takes the block diagonal form 
\beq\label{RegH}
\begin{split}
H= \sum_{lm} \int \frac{dk}{4 \pi }  \frac{dk'}{4 \pi }  \Bigl[&A_{lm}(k, k') a_{k lm}^{\dagger}a_{k'lm} + B_{lm}(k, k') a^{\dagger}_{k lm}a^{\dagger}_{k'l,-m} + \\ 
&\tilde{A}_{lm}(k, k')\tilde a_{k lm}^{\dagger}\tilde a_{k'lm}+\tilde{B}_{lm}(k, k')\tilde a^{\dagger}_{k lm}\tilde a^{\dagger}_{k'l,-m}  + c.c.\Bigr] \ .
\end{split}
\eeq 
Here for a stationary slicing
\beq
A_{lm}(k, k') = \frac{1}{2kk'} (\phi_{klm}, i\partial_t \phi_{k'lm})
\eeq 
and
\beq
B_{lm}(k, k') = \frac{1}{2kk'} ( \phi_{klm}, i\partial_t \phi^*_{k'l,-m})\ ,
\eeq 
 $\tilde{A}_{lm}(k, k')$ and $\tilde{B}_{lm}(k, k')$ are defined similarly for the in-modes, and $c.c.$ denotes conjugation that doesn't change operator ordering.\footnote{Reordering then yields a normal ordering constant.}
 Note that the Hamiltonian does not contain mixing terms between $\mathcal{H}_{in}$ and $\mathcal{H}^R_{up}$
due to the orthogonality of the basis modes. This is particularly clear when writing the Hamiltonian in terms of the Bogolubov coefficients. Using the expansion \eqref{Bogreln} and relation \eqref{tderiv}, the mixing terms reduce to inner products between orthogonal energy eigenmodes. The remaining nonzero terms of the regular Hamiltonian \eqref{RegH} are then characterized by the functions
\beq
A_{lm}(k, k') =\frac{1}{2kk'} \int d\omega  4\pi \omega^2 (\beta^{+*}_{k\omega l} \beta^+_{k'\omega l}+ \beta^{-*}_{k\omega l} \beta^-_{k'\omega l} - \hat \beta^{+*}_{k\omega l} \hat\beta^+_{k'\omega l}- \hat \beta^{-*}_{k\omega l} \hat \beta^-_{k'\omega l})\ ,
\eeq 
and
\beq
B_{lm}(k, k') =  \frac{(-1)^{-m}}{2kk'}\int d\omega4\pi \omega^2 (\beta^{+*}_{k\omega l} \beta^{-*}_{k'\omega l}+ \beta^{-*}_{k\omega l} \beta^{+*}_{k'\omega l} - \hat \beta^{+*}_{k\omega l} \hat\beta^{-*}_{k'\omega l}- \hat \beta^{-*}_{k\omega l} \hat \beta^{+*}_{k'\omega l})
\eeq 
where $\tilde{A}_{lm}(k, k')$ and $\tilde{B}_{lm}(k, k')$ are defined similarly, but with only one set of coefficients $\tilde\beta^+$ and $\tilde{\beta}^-$.  With specific choices of the mode functions the Bogolubov coefficients as well as the coefficients $A$ and $B$ can in principle be calculated.

Given the hamiltonian \eqref{RegH},  the evolution is in principle well defined.  In practice, evolution in such a regular description is more complicated than in the singular description in terms of energy eigenmodes.  It has of course been of interest to establish that there {\it is} a regular description, as well as to understand aspects of its behavior.  We will also explore its relation to the description using energy eigenmodes, and how properties of the evolving wavefunction can consequently be inferred.

\section{Evolution for dynamic black holes and the ``Hawking state"}
\label{BHevol}

In this section, we will extend the preceding discussion to consider evolution of quantum matter on a general time-dependent, spherically-symmetric BH background, corresponding for example to a BH that forms from collapse of a massive body, and will discuss some properties of the corresponding quantum state.  

\subsection{Geometry}

\begin{figure}[!hbtp] \begin{center}
\includegraphics[width=9cm]{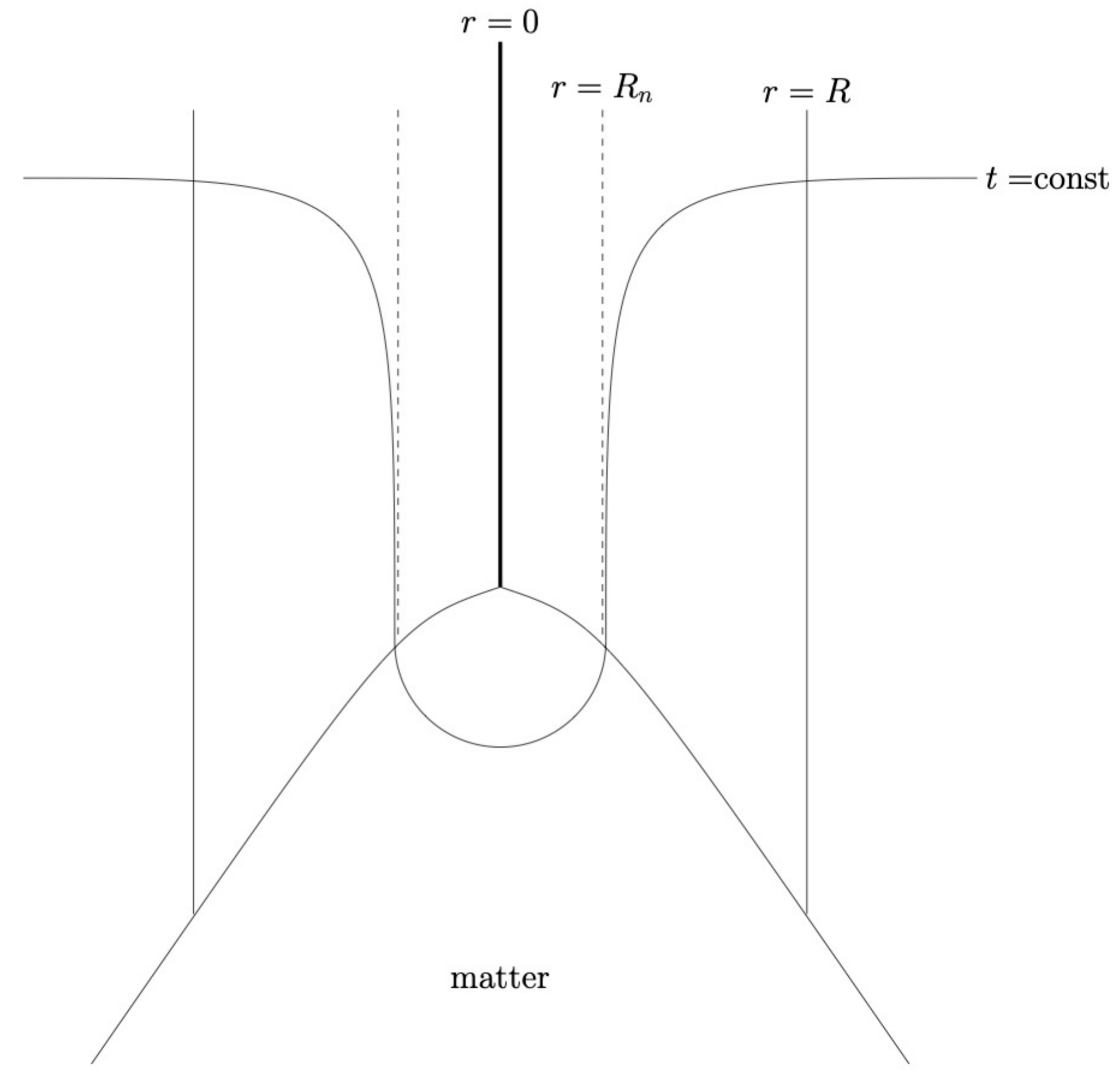}
\end{center}
\caption{The geometry of a black hole formed from collapse, in an Eddington-Finkelstein diagram.  Also shown is a slice that behaves like a nice slice in the vacuum region, which is then extended through  the collapsing region to complete it to a Cauchy slice.}
\label{Collapse}
\end{figure}

Specifically, consider the general metric
\beq\label{gmet}
ds^2 = -f(x^+,r) dx^{+2} + g(x^+,r) dx^+ dr + r^2 d\Omega^2\ ,
\eeq
where the null ingoing coordinate $x^+$ can be chose so that  $g(x^+,r)\rightarrow 2$ at $r\rightarrow\infty$.  This could represent the metric of a general collapsing matter distribution, as shown in Fig.~\ref{Collapse}; a specific case is the ingoing Vaidya solution with
\beq \label{fVaid}
f(x^+,r)= 1-\frac{2M(x^+)}{r}\quad,\quad g(x^+,r)=2\ ,
\eeq
with an ingoing mass function $M(x^+)$.

If we wish to provide a slicing of a collapsing BH spacetime such as shown in Fig.~\ref{Collapse} by Cauchy slices, those slices need to avoid the singularity at $r=0$.  
We assume that slices in the far past, when matter is dilute, are increasingly close to Minkowski time slices.  Later slices can remain Cauchy
 if they have ``nice" behavior, with a minimal radius $R_n$. These slices may then be closed in the region prior to the singularity, as illustrated in the figure.  If we suppose that these slices asymptote to Minkowski time slices, the portion in the vacuum region needs to advance  into the future with advancing time.

There are different ways to accomplish this, corresponding to different choices of coordinates or gauge, which arise from different choices of the functions ${\cal X}^+(t,x)$ and ${\cal X}^r(t,x)$ in \eqref{xrparam}.  For example, these could be chosen so that the spatial metric $q_{xx}$ on the slices of constant time $t$ is time-dependent in the post-formation vacuum region, and might specifically undergo ``stretching" as described in \cite{Mathinforev}.  However, the description of the state in the post-formation region of interest is simpler if we instead use stationary slices, as in \eqref{statgen}, in this region.  Since the distance along the slices back to $r=0$ must increase with time, these slices must undergo ``stretching" in the early region, for example within or near the infalling matter region.  In this section we will not focus on the latter behavior since it appears most relevant to the description of the ``early" part of the BH state, and our interest will be in the part of the state corresponding to Hawking radiation and internal excitations at later times.

\subsection{States}

Given a slicing, we will define the ``Hawking state" as the time-dependent state $|\psi(t)\rangle_H$ that arises from evolving the matter vacuum $|0\rangle$ at $t=-\infty$ to a future time $t$.  Here, we focus on ``spectator" matter that is different from the matter forming the BH.  We expect the Hawking state to be given by the expression
\beq\label{Hevol}
|\psi(t)\rangle_H = T e^{ -i \int_{-\infty}^tdt H_\xi}|0\rangle\ ,
\eeq
with a hamiltonian as discussed in Sec.~\ref{HandP}.  This expression will implicitly depend on the choices of coordinates and mode bases, described in Sec.~\ref{Schsec}, used to define the picture.

The full time-dependent Hawking state \eqref{Hevol} can be rather complicated, in  part due to excitations created during the time-dependent BH formation phase.  A simpler state that is sometimes considered is the Unruh state\cite{Unru}, which can be defined by working with the extended vacuum Schwarzschild solution, and evolving the vacuum defined with respect to the Kruskal coordinate $X^-$ at the past horizon forward in time, via a similar procedure.

Indeed, if we compare the Hawking and Unruh states on a time slice that meets the horizon just after the transition to vacuum, such as shown in Fig.~\ref{Collapse}, we expect that they differ in some of the excitations escaping to infinity or falling into the BH.  But, if we consider a much later slice, these excitations will have reached the asymptotic region, or reached the deep interior, either near $r=0$, or with the kind of slicing we have described, the part of the slice at $r=R_n$.
In contrast, the excitations being emitted from or falling into the BH near that later time are expected to be determined by the local short-distance structure of the state, which is the same for the Hawking and Unruh states.

In fact, we expect these statements to extend to a more general state that behaves like vacuum near the horizon at short distances -- the subsequent long time behavior is governed by this vacuum-like structure near the horizon.  Specifically, we expect that any regular state has the same long-time behavior, once excitations that correspond to initial differences between states have escaped to near infinity or to  the BH deep interior.  The evolution of such a state can be examined at a more explicit level.  

\subsection{Evolution}

 To describe the evolution, first one makes a choice of slicing and coordinates along those slices, given by \eqref{xrparam}, or equivalently by specifying $t(x^+,r)$ and spatial coordinate $x(x^+,r)$, which we assume to be regular across the horizon.  Next, choose a time $t_0$ which is taken so that the corresponding slice meets the horizon to the future of the collapsing matter as in Fig.~\ref{Collapse}.  To begin, we wish to characterize what it means to be vacuum-like at this time near the horizon.
 
The vacuum-like structure can be characterizing by using the local relation to flat Minkowski geometry.  In the vacuum region, the metric \eqref{gmet}, \eqref{fVaid} can be written in terms of Kruskal coordinates, defined for example  in $D=4$ by 
\beq\label{Krudef}
X^\pm = \pm 2R e^{\pm x^\pm/2R}\ ,
\eeq
with corresponding extension across the horizon.\footnote{For more discussion of these coordinates and the Rindler limit, see Appendix~\ref{Kapp}.} The vacuum metric then becomes
\beq \label{Krmet}
ds^2 = -\frac{R}{r}e^{1-\frac{r}{R}}dX^+ dX^- + r^2 d\Omega^2\ .
\eeq
The near horizon limit $|r-R|\ll R$ gives a ``Rindler region"\cite{GiLi3} in which the metric is locally $M^2 \times S^2$,
\beq\label{prodmet}
ds^2 \approx -dX^+ dX^- + R^2 d\Omega^2\ ,
\eeq
 with corresponding 2d spacetime coordinates defined by $X^\pm = T\pm X$.  In this near horizon limit and in these coordinates, a slice whose slice function $S(r)$ only varies on scales $\sim R$ will meet the horizon as a straight line.  The regular basis $(\phi_{klm},\pi_{klm})$, $(\tilde \phi_{klm},\tilde \pi_{klm})$ can then be chosen so that at high $k$, the corresponding solutions behave as
 \beq\label{KruskalModes}
 u_{kl}\approx e^{ikX-i\omega_k T}\quad , \quad \tilde u_{kl}\approx e^{-ikX-i\omega_k T}
 \eeq
 in the Rindler region,\footnote{A more careful treatment requires localization of the modes of the basis.  This can be done by constructing wavepackets, for example as described in \cite{Hawk}\cite{GiNe}\cite{SE2d}.} with $\omega_k^2-k^2=[l(l+1)+1]/R^2$.  Then, with the field expansion \eqref{Fexp}, a state which is locally Minkowski is one satisfying $a_{klm}|\psi\rangle=\tilde a_{klm}|\psi\rangle=0$ for the operators associated to $k\gg 1/R$, $l\gg 1$ modes in this region.  Of course, modes with $k\lesssim 1/R$ can be excited in such a state.
 
The evolution of such a state is in general governed by the regular expression \eqref{RegH} for the hamiltonian.  In the Rindler region this simplifies to give Minkowskian evolution, but this receives nontrivial corrections as excitations reach $|r-R|\sim R$.  This nontrivial behavior causes excitations of the local vacuum.  The details of this evolution depend on the detailed structure of the hamiltonian \eqref{RegH} and its corresponding evolution operator, which can be somewhat complicated.  However, one thing that we do immediately learn in this description is that the transition to excited states takes place on scales with $|r-R|\sim R$, as was also found in 2d\cite{SEHS,SE2d}.  This supports previous arguments\cite{SGBoltz} (see also the earlier related arguments \cite{Unru-origin,Full,Bard}) that Hawking radiation ``is produced" in a black hole atmosphere at these scales, as opposed to at ultrashort distances.  

We have argued that the hamiltonian \eqref{RegH} gives a regular description of evolution on slices avoiding the singularity, but one with complicating features.  
To learn more about the state one would like a simpler description of this evolution.  This is achieved by going to the energy eigenbasis of Sec.~\ref{EnergyEmodes}. While this description is inherently singular, as we have seen, it does furnish an effective way to more simply describe the evolution, due to the simplicity of the hamiltonian  \eqref{EMham} in this basis.  To describe the latter evolution, we must first rewrite a regular state $|\psi\rangle$ in this basis.  This in principle follows from the Bogolubov transformation \eqref{Bogreln}, which is however also complicated.  But, as discussed, the long-time behavior is expected to follow from the local Minkowski structure near the horizon.

In particular, in the high-$k$  Rindler region limit, the regular modes can be chosen to simplify to the form $\exp\{-ikX^\pm\}$.  The locally right-moving modes $\exp\{-ikX^-\}$ there are related to the local (approximate) energy eigenmodes $\exp\{-i\omega x^-\}$, $\exp\{-i\omega \hat x^-\}$ by the same relation that relates Minkowski to Rindler modes, as seen from the coordinate transformation \eqref{Krudef}.  Specifically, from this we find that the combinations
\beq
e^{-i\omega x^-} + e^{-2\pi R\omega} e^{i\omega \hat x^-}\quad ,\quad e^{-i\omega \hat x^-} + e^{-2\pi R\omega} e^{i\omega  x^-}
\eeq
are analytic in the lower half complex $X^-$ plane, and thus correspond to positive frequency modes in $X^-$.  Then the corresponding operators
\beq
b_\omega - e^{-2\pi R\omega}\hat b^\dagger_\omega\quad,\quad \hat b_\omega - e^{-2\pi R\omega} b^\dagger_\omega
\eeq
correspond to Minkowski annihilation operators, which should annihilate the state for large $\omega$.  
Thus, for these high-wavenumber modes, the regular state has local description
\beq\label{statereln}
|\psi\rangle \sim\sum_{\{n_\omega\}} e^{-2\pi R\int d\omega \omega n_\omega }| \{\hat n_\omega\} \rangle |\{n_\omega\}\rangle\ 
\eeq
in terms of occupation number eigenstates for the $\hat b_\omega$ and $b_\omega$, and analogously for higher $D$.
Just as with Minkowski space, this description is inherently singular.  However, it is useful, and may for example be regulated with an appropriate short distance cutoff.

This description of the state is useful because it provides an effective intermediary to relate evolution of modes near the horizon to corresponding asymptotic modes.\footnote{This can be seen even more explicitly in the two-dimensional example\cite{SE2d}, which avoids complications such as reflection/transmission near the horizon.}  Consider  excited modes in the expression \eqref{statereln}.\footnote{Again, a more precise version of this argument would use wavepackets to localize modes in both position and frequency.}  These are evolved by the simple hamiltonian \eqref{EMham}, which vanishes for paired excitations, and a near-horizon wavepacket of such modes will evolve into a future wavepacket of the same modes.  In the case of the ``up" modes, associated to the $b_{\omega l m}$, the wavepackets will have an outgoing piece at infinity, with magnitude given in terms of the transmission coefficient for the effective potential $fV$ of \eqref{diffeq}, and a reflected part that enters the BH.

Once we have used these modes as {\it intermediaries} to simplify the description of the state, we can alternately convert back into a regular mode basis.  In fact, in the asymptotic region, we expect the energy eigenmodes $u_{\omega l}$ to equate to corresponding regular modes, up to the factor of the transmission coefficients, since they are both governed by free Minkowski evolution.   Specifically, asymptotically we expect these modes to take the form of flat space modes, with wavepackets that are linear superpositions of
\beq
\phi_{\omega l m} \sim T_{\omega l}\, j_l(kr) e^{-i\omega t}\frac{Y_{lm}}{r^{D/2-1}}\ ,
\eeq
where $T_{\omega l}$ is the relevant transmission coefficient.  These can be regarded either as energy eigenmodes or as regular modes, in this region.
 In short, we convert to the energy eigenmode basis to simplify the evolution out to the asymptotic region, and then convert back to a regular basis using this relation between modes.
In this fashion, the intermediaries provide a simple way to characterize the result of evolution of the regular expression for the state with the regular hamiltonian \eqref{RegH}.   It has a thermal spectrum at the expected temperature, {\it e.g.}  $T=1/4\pi R$ for $D=4$, following from the form of \eqref{statereln}.  It also has the expected pairing and entanglement between  quanta of Hawking radiation, and corresponding internal excitations of the BH, implied by the pairing in  \eqref{statereln}, along with the transmission factors.  Again, this will be the generic long-time behavior, after transitory excitations, of states that are regular at the horizon.

We emphasize that in this discussion, the singular energy eigenmodes are {\it only} used as intermediary tool, and are not taken as part of a literal fundamental description of the state.  
This differs from a significant part of the literature, in which the singular energy eigenmodes are sometimes viewed as playing a more fundamental role; here, we stress, they are merely a convenient basis for some purposes.  In practice, one way to work with a description of the state in terms of them is to introduce cutoffs in the description.  
But not  regarding these modes as fundamental  avoids the potential pathologies that arise if these modes are regarded as true physical excitations.

\subsection{Internal evolution: nice slices and freezing}

Evolution via a local quantum field theory hamiltonian, such as \eqref{LapseForm} or \eqref{StressTensorForm}, is expected to plausibly give a good approximate description of the complete physical evolution of excitations outside but near a BH, although one expects the need for important, but possibly small,  corrections to ultimately restore unitarity\cite{SGmodels,BHQIUE,NVNL,NVU,BHQU}.  On the other hand, one expects that the evolution of excitations inside the BH is likely to ultimately receive large corrections.  Nonetheless, it seems of interest to better understand the leading field theory description of the internal evolution, as background and preparation for understanding its possible modifications.

Such evolution can be described on a family of Cauchy slices.  As was noted in Sec.~\ref{Slicing}, slices that reach $r=0$ are not Cauchy, and so a description on such slices must be supplemented by additional dynamics ``at $r=0$."  But, evolution may be considered on a family of slices that avoid $r=0$, and in particular on a family of nice slices that asymptote to a minimal radius $r=R_n$.  We will describe some features of evolution on these slices.  Our focus will be on the vacuum region of the BH, and we will consider stationary slices, as specified in \eqref{statgen} or \eqref{statS}, with a slice function chosen to asymptote to $R_n$. 

As we have noted, one also needs choices of spatial coordinate and modes to describe evolution of the state.  The use of $r$ as a spatial coordinate on the constant-$t$ slices leads to a coordinate system $(t,r)$ that degenerates at $r=R_n$.  This means it is preferable to use a more general spatial coordinate, $x(t,r)$.  

The choice of a ``stationary" coordinate $x(r)$ (say, with $x\rightarrow -\infty$ as $r\rightarrow R_n$) results in a nonzero shift $N^x$ at large $t$, as the slices accumulate at $r=R_n$, although the lapse $N$ vanishes in this limit.  This implies a nontrivial contribution to the hamiltonian in this region, as seen for example from \eqref{ADMH}.  This may be alternately understood by considering the form of the wavefunction solutions.  For example, in such coordinates, the solutions \eqref{Eefcns} take the form 
\beq
e^{-i\omega t} U_{\omega lm}(x,\Omega)\ ,
\eeq
and thus continue to have nontrivial time evolution as $t\rightarrow\infty$.

On the other hand, it is clear from the accumulation of slices at $r=R_n$ that the evolution of a state can be described as freezing\cite{QBHB}\cite{BHQIUE} at this radius, as $t\rightarrow \infty$.  This is best described with a choice of {\it non-stationary} coordinate along the slices.  This choice can be specified through a more general relation $r(t,x)$, as in \eqref{xrparam}.  Then, from the ADM form \eqref{ADMmet} of the metric, we find $N$ to be unchanged from \eqref{ADMvar}, and
\beq\label{Xmmet}
q_{xx}=r^{\prime 2} q_{rr}\quad ,\quad N^x = \frac{\dot r + N^r}{r'}\ ,
\eeq
with $r'=(\partial r/\partial x)_t$, $\dot r= (\partial r/\partial t)_x$.  With such coordinates, the lapse, shift, and spatial metric are now explicitly time-dependent, also resulting in an explicitly time-dependent expression for the hamiltonian \eqref{ADMH} or \eqref{LapseForm}.  

One way to exhibit the freezing behavior is if $x\rightarrow x^+$ as $r\rightarrow R_n$, which leads to both a vanishing lapse and shift as $r\rightarrow R_n$, and so vanishing hamiltonian density there.  An example\cite{SE2d} is the coordinate $x=x^+ + g(r)$, with $g$ vanishing as $r\rightarrow R_n$, although more generally we might  instead like to use a time-dependent function $g(t,r)$ so that the coordinate $x$ matches $r$ asymptotically as $r\rightarrow\infty$, which is achieved if $g(r,t)\approx -t$ in this limit.  

The freezing simplifies the description of the internal part of the state, since it no longer evolves, and thus gives one way of simply describing BH internal states in terms of this static appearance, in this approximation.  

In these coordinates and this picture, the slices exhibit stretching behavior, rather than translating under a shift in $t$.  For example, the distance along a slice from a given fixed $x$ corresponding to a point near $r=R_n$ increases linearly in $t$.  This arises from the time dependence of $q_{xx}$ from \eqref{Xmmet}, and may for example be concentrated in the vicinity of the horizon depending on the specific choice of $r(t,x)$.  
 The explicit time dependence of the metric and hamiltonian in this gauge  and picture introduces additional subtleties which we defer to future work, but which we expect may be resolved by connecting back to the underlying stationary description, also in analogy to \cite{MuOe}.

\section{Extensions: interactions, generalizing asymptotics, AdS and connection to $1/N$}

The discussion of the bulk of this paper has been of a noninteracting theory such as \eqref{Sact}, with flat asymptotic geometry, but it is expected that the quantum description extends both to interacting theories, and to more general, {\it e.g.} AdS, asymptotics, where there is also a connection with the large $N$ limit of the AdS/CFT correspondence.

\subsection{Evolution in interacting theories}

The extension to interacting theories, and theories with higher-spin matter, is evident; beginning with a generalization of the action \eqref{Sact} to incorporate interactions and/or higher spin, 
 the canonical approach yields a hamiltonian of the quadratic  form \eqref{ADMH}, together with additional interaction terms.  Canonical quantization proceeds from there via the canonical commutators, \eqref{CCRs}, or their higher-spin generalization.  With a choice of basis of regular mode functions, and expansion in ladder operators analogous to \eqref{Fexp}, this results in a hamiltonian of the form \eqref{RegH}, together with higher-order terms describing the interactions.  While, as expected, this can result in more complicated dynamics, the evolution of the state, {\it e.g.} as in \eqref{Hevol}, is in-principle concretely defined, modulo usual issues of renormalization, {\it etc.} 

This observation illustrates two points.  The first is that the present methods extend beyond Hawking's original derivation\cite{Hawk}, which relied on use of the free propagating mode functions and so did not easily incorporate interactions.  In evolution such as \eqref{Hevol}, the state continuously evolves in $t$ according to the structure of the hamiltonian, without direct dependence on having solutions.  It is important to have such a generalization, to treat Hawking radiation in interacting theories.  The second point is that the specific choice of the mode functions is less important with this additional context.  That is because a particular choice of modes may be motivated so to simplify evolution in the non-interacting case; a specific example is that of energy eigenstates, where the hamiltonian greatly simplified, to \eqref{EMham}.  However, once interactions are included, such simplifications are lost.  In the interacting theory, it appears that different choices of regular mode bases won't make significant difference in the practical difficulty of describing the evolution of the state, and so fairly general choices can be considered.

This discussion also extends to include gravitational perturbations, and their couplings to perturbations of other fields.  The full fluctuating metric may be expanded $\tilde g_{\mu\nu} = g_{\mu\nu}+ \kappa_D h_{\mu\nu}$, with $g_{\mu\nu}$ the metric of the BH background and $\kappa_D^2 =32\pi G_D$.  Then, the action and hamiltonian for the metric fluctuation $h_{\mu\nu}$ have a leading quadratic term similar to that for scalars (but also requiring gauge fixing), and interaction terms between $h_{\mu\nu}$ and the other fields, as well as self interactions of $h_{\mu\nu}$ at higher orders in the expansion in $\kappa_D$.  By the steps just outlined, these interactions lead to an interacting hamiltonian generalizing \eqref{ADMH},  \eqref{RegH}, which may be treated by the same methods, to determine the evolution of the state on a chosen set of slices.  

\subsection{Generalizing asymptotics, and AdS/Schwarzschild}

Our main discussion has focussed on asymptotically flat spacetime.  However, we expect it to extend to BHs with other asymptotics, using a slicing analogous to that described in Sec.~\ref{Slicesec} that is taken to similarly extend to the BH interior.  Of course with general asymptotics, we may need to confront further subtleties associated with lack of a Killing vector corresponding to time translations.  A prominent case with such a Killing vector is that of BHs in AdS.  For example, the $D$-dimensional AdS/Schwarzschild solution for mass $M$ takes the form \eqref{Schmet}, with
\beq\label{AdSSch}
f(r) = 1 + \frac{r^2}{R_\Lambda^2} - \left(\frac{R_0}{r}\right)^{D-3}\ ;
\eeq
here $R_\Lambda$ is the AdS radius, and 
\beq
R_0^{D-3} = \frac{16\pi G_D M}{(D-2)A_{D-2}}
\eeq
with $A_{D-2}$ the area of the unit sphere.  One may alternately use an ingoing null coordinate $x^+$ to rewrite the metric in the form \eqref{EFmet}, with $f$ given by \eqref{AdSSch}, and exhibit both exterior and interior of the BH.  A trans-horizon slicing analogous to those of \eqref{sliceparam}-\eqref{statS} may then be used to describe interior and exterior, and quantization may be performed analogous to the discussion of Sec.~\ref{Schsec}, resulting in  hamiltonian evolution analogous to that of \eqref{Hevol}.  In particular, we expect the methods of this paper to yield long-time thermal behavior for a large class of regular initial states.  These would evolve similarly to the description of  Sec.~\ref{BHevol}, with the additional feature that the AdS asymptotics behave like a reflecting cavity, and so Hawking excitations are expected to be reflected back towards the BH.  

Specifically, we may describe the interacting hamiltonian and evolution perturbatively in Newton's constant $G_D$.  The leading order evolution (also expanding in other couplings, if present) is hamiltonian evolution of free fields, including the graviton perturbations, by a hamiltonian analogous to \eqref{ADMH},  \eqref{RegH}, on the AdS/Schwarzschild background.  Couplings to gravitational perturbations, backreaction, dressing, {\it etc.} then arise at higher order in $\kappa_D$.

\subsection{AdS/CFT and large-$N$ description}

Gravitational dynamics in AdS is conjectured to be equivalent to that of a ``boundary"  CFT\cite{Mald}.  In this context, it is interesting to explore the possible relation between the perturbative dynamics we have outlined, and the dynamics of the CFT.  We focus on the ``classic" example of AdS/CFT, with $AdS_5\times S^5$ dynamics conjectured to be dual to ${\cal N}=4$ SU(N) super Yang-Mills on $S^3\times \mathbb{R}$.  The parameters are related by $(R_\Lambda/l_{10})^4 \sim  N$, where  $l_{10}^8 \sim G_{10}$ gives the ten-dimensional Planck length, and formulas here correctly include parameters but neglect numerical $\calo(1)$ factors.  Black holes with horizon radii $R\ll R_\Lambda$ are expected to be ten-dimensional localized objects in $AdS_5\times S^5$; those with $R\gtrsim R_\Lambda$ are expected to behave as five-dimensional BHs \eqref{Schmet}, \eqref{AdSSch} that are uniform on $S^5$, and these two cases are expected to be connected by a Gregory-Laflamme transition\cite{GrLa}.  The transition radius $R\sim R_\Lambda$ corresponds to a mass threshold $M\sim N^2/R_\Lambda$.  

Thus, the case of AdS BHs, here with $D=5$, is strictly speaking only valid above this threshold.  But in this case, the leading order dynamics, in a QFT description of the bulk, is expected to be given by a free hamiltonian like \eqref{ADMH},  \eqref{RegH}.  And, this is expected to receive perturbative corrections, order-by-order in the gravitational coupling.  Given the relationship between parameters, and the relation $G_{10}\sim R_\Lambda^5 G_5$, this corresponds to an expansion in $\kappa_5\sim R^{3/2}_\Lambda/N$.  

As is known, this connects gravitational perturbation theory with the large $N$ limit and $1/N$ expansion.  A first question is what is held fixed as $N$ is taken to be large.  We will focus on the large $N$ limit with AdS radius $R_\Lambda$ held fixed, when then corresponds to the limit of $G_{10}$ or $G_5$ becoming small.  If we wish to consider BH states, then the preceding scaling tells us that we need to consider states whose energy scales up as $N^2$.  However, if we consider for example fixing the temperature, {\it e.g.} as in \cite{ScWi}, that is equivalent to holding the radius of the BH fixed and so the geometry  \eqref{Schmet}, \eqref{AdSSch} is unchanged as $N$ increases.

The hamiltonian that describes BH excitations in the large $N$ limit is then of the form described in the preceding subsections.  Specifically, a candidate infinite-$N$ hamiltonian $H_\infty$ can be found by choosing a slicing for AdS-Schwarzschild like those described in Sec.~\ref{Slicesec}, and specifically avoiding the singularity, and then deriving the corresponding  hamiltonian \eqref{ADMH} (or \eqref{RegH}) which is quadratic in each of the field perturbations that propogate on AdS.  

Moreover, $1/N$ corrections to $H_\infty$ correspond to the bulk interaction terms between these perturbations that arise in the perturbative expansion in $\kappa_5$. 

Questions have recently been raised about the existence of a large $N$ description of BHs in AdS/CFT in \cite{WittQFT,Wittcross},\cite{ScWi}.  The present construction appears to begin to provide answers, and specifically to address the statement\cite{ScWi} that the literature doesn't contain a proposal for the hamiltonian for a BH in the large $N$ limit.

Of course, what {\it is} expected to be true is the statement that the perturbative hamiltonian that is found this way does not give a complete description of the BH dynamics:  specifically, there are good reasons to believe that this perturbative hamiltonian does not ultimately lead to unitary evolution, and one piece of that evidence is the fact that $H_\infty$ can describe an infinite number of BH states.  

For this reason, we expect the complete gravitational hamiltonian will be a corrected version of the perturbative hamiltonian $H_{pert}$ we have just described:
\beq
H = H_{pert} + \Delta H\ .
\eeq
We would obviously like to understand the structure of $\Delta H$, and what phenomena it encodes; another pertinent question is its dependence on $N$.

It has previously been argued\cite{NVNL,NVNLT,NVU,BHQU} in the case of flat asymptotics that $\Delta H$ has two important pieces: a piece $\Delta H_I$ containing interactions between the BH states and the BH's surroundings, necessary to transfer information or entanglement from the BH, and a piece $\Delta H_{BH}$ modifying the internal dynamics of the BH states.  Simple forms of $\Delta H_I$ have been parameterized\cite{NVU}, and it is plausible that these corrections are small, even nonperturbatively so, in $N$.  On the other hand, we might expect $H_{pert}$ to receive large internal corrections in $\Delta H_{BH}$  corresponding to corrections to dynamics in the strong-curvature regime at the core of the BH.  
This is expected to be necessary, for example, to ensure a finite number of internal BH states.
It is plausible that these corrections also yield chaotic internal behavior.  Their dependence on $N$ is less clear, but it is quite plausible that there are also important nonperturbative contributions here.  A possible role for ensemble averages, like discussed in \cite{ScWi}, is also less clear, unless the corrections to $H_{pert}$ for example arise from baby universe emission\cite{Cole,GiSt,MaMa,GiTu,SGthm,HTY}.

In such a plausible picture for $\Delta H$, new chaotic dynamics is only associated with the deep interior dynamics of the BH; evolution in the near-horizon regime, both inside and outside the BH, may be close to that of LQFT, with only relatively small corrections that, for example, an infalling observer would perceive as innocuous.  Further discussion of this picture, which may also lead to observational effects for BH observations\cite{SGObs,SGObs2,SGAstro}\cite{BHQU}, is given in the works cited above.  Needless to say, it would be very interesting if such effects could be understood from, or even derived from, the AdS/CFT correspondence. Or, perhaps, they have a different explanation.

\vskip.3in
\noindent{\bf Acknowledgements} 

This material is based upon work supported in part by the U.S. Department of Energy, Office of Science, under Award Number {DE-SC}0011702, and by Heising-Simons Foundation grant \#2021-2819.   We thank T. Jacobson and S. Mathur for useful discussions.

\appendix
\section{Radial Equation and Heun Function for D=4}
\label{Happend}

A more detailed picture of energy eigenmodes can be gained by further examining the equation of motion either in the form of \eqref{rxpeq} or \eqref{diffeq}. We will focus our discussion on the case of Eddington--Finkelstein coordinates, equation \eqref{rxpeq}, to connect to the analysis in Sec.~\ref{EnergyEmodes}. The Schwarzschild coordinate solutions of \eqref{diffeq} will be related by the transformation \eqref{rescm} to the solutions described in this appendix.

The ansatz $u(x^+, r) = r e^{-i \omega x^+} y(r)$ can be used to rewrite the differential equation \eqref{rxpeq} as 
\beq\label{Heun}
y''+p_0(r)y'+ p_1(r)y = 0\ ,
\eeq
where the prime denotes derivatives with respect to $r$,
\beq
p_0(r) = \frac{-2 i \omega r^{D-2}+2r^{D-3}+(D-5)R^{D-3}}{r(r^{D-3}-R^{D-3})},
\eeq
and
\beq
p_1(r)=\frac{-2 i \omega r^{D-2}-\left[ l(l+D-3)  + (D-2)(D-4)/4\right]r^{D-3} + \left[(D-3)-(D/2-1)^2\right] R^{D-3}}
{r^2(r^{D-3}-R^{D-3})}.
\eeq
It can be seen that in arbitrary dimension the differential equation has regular singular points at~$0$ and $R e^{2 \pi i n/ (D-3)}$ for integers $n = 0, 1,..., D-4$, and an irregular singular point at infinity. The solutions of \eqref{Heun} are unknown in arbitrary dimension, but by defining a rescaled spatial variable $x = r/R$, the $D=4$ equation can be rewritten  as
\beq\label{CHeun}
y''+\Big(\frac{1}{x}+\frac{1-2 i \omega R}{x-1}-2 i \omega R\Big)y'+ \Big(\frac{-2i \omega R x-l(l+1)}{x(x-1)}\Big)y = 0\ ,
\eeq
which is the confluent Heun equation;\footnote{Some references discussing this equation and its relevance to BHs are \cite{Leav,Fizi1,Fizi2,PhPe}.}  the case of general $D$ thus represents a generalization of the confluent Heun equation.  A standard form for the confluent Heun equation is
\beq\label{CHeunGen}
y''+\Big(\frac{\gamma}{z}+\frac{\delta}{z-1}+\epsilon\Big)y'+ \Big(\frac{\alpha z-q}{z(z-1)}\Big)y = 0\ ,
\eeq
and we denote the solutions to \eqref{CHeunGen} that satisfy the  regularity condition $y= 1$ at the singular point $z=0$ as 
as $\mathrm{HC}[q, \alpha, \gamma, \delta, \epsilon, z]$.  This confluent Heun function is  implemented in Mathematica as $\mathrm{HeunC}$, with  parameters as in \eqref{CHeunGen}. 

The incoming and outgoing modes of interest can be specified by their behavior in the vicinity of the horizon. The three solutions  used to define the basis outlined in 
Sec.~\ref{EnergyEmodes} are found as follows.  First, we let $x=1-z$ in \eqref{CHeun}, and compare the resulting equation to \eqref{CHeunGen}; this gives the solution $\tilde u_{\omega l}$ regular at the horizon.  For the up modes, we then substitute $y\rightarrow (-z)^{2i\omega R} y$ into the resulting equation, which gives the confluent Heun equation with different coefficients.  The inside mode is found in the same fashion.  This results in the the following explicit solutions.

\begin{itemize}
\item $\tilde{u}_{\omega l} = r \, \mathrm{HC}[l(l+1)+2 i \omega R, 2 i \omega R, -2 i \omega R + 1, 1, 2 i \omega R, 1-r/R]$ is the incoming mode, which is regular at the horizon. From \eqref{rescm}, one sees that the corresponding function $g_{\omega l}$ is not regular.

\item $u_{\omega l} = r(r/R-1)^{2i\omega R} \, \mathrm{HC}[l(l + 1) - 4 \omega^2R^2, 2 i \omega R - 4 \omega^2R^2, 2 i \omega R + 1, 1, 2 i \omega R, 1-r/R]$ is the up mode solution, which is not regular at the horizon, and gives the outgoing Hawking mode.

\item $\hat{u}_{\omega l}^* = r(1-r/R)^{2i\omega R} \, \mathrm{HC}[l(l + 1) - 4 \omega^2R^2, 2 i \omega R - 4 \omega^2R^2, 2 i \omega R + 1, 1, 2 i \omega R,  1-r/R]$ is the inside mode solution. It is also not regular at the horizon, and corresponds to the internally trapped Hawking partner mode. It is defined inside the horizon, for $0<r/R<1$.
\end{itemize}
The analysis in Sec.~\ref{EnergyEmodes} uses the asymptotic behavior  $e^{-i\omega x^+}$ in the far past of the incoming solution near infinity, and  $e^{-i\omega x^-}$ of the outgoing solution near the horizon, as well as $e^{-i\omega \hat{x}^-}$ of the corresponding partner  inside the horizon.

\section{Kruskal coordinates and Rindler region}
\label{Kapp}

While the coordinates $(x^+,r)$ are useful for exhibiting the time translation symmetry, Kruskal coordinates $X^\pm$ are useful for exhibiting the Minkowski-like structure of the near-horizon Rindler region.  Since the time translation symmetry becomes a scaling (boost) symmetry in these coordinates, this symmetry becomes less transparent in the equations of motion in these coordinates.  In this appendix, we collect some basic results on this Kruskal description, in the example of $D=4$.

As was described in the main text, the Kruskal coordinates are related to the Eddington--Finkelstein coordinates by
\beq\label{Kcoorddef}
X^{\pm} = \pm 2R e^{\pm{ x^{\pm}}/{2R}}\ ,
\eeq
with a continuation across the horizon in terms of $\hat x^-$ such that the vacuum metric, given in \eqref{Krmet},  is regular at the horizon.  
In the Rindler region $|r-R|\ll R$, or $|X^+X^-|\ll R^2$, the metric is well approximated as that of $M^2\times S^2$, as seen in \eqref{prodmet}, with local Minkowski spacetime coordinates defined by $X^\pm=T\pm X$.  The radial coordinate is related to Kruskal coordinates by
\beq
X^+X^-=4R^2 \left(1-\frac{r}{R}\right) e^{r/R-1}\ ,
\eeq
or
\beq\label{rtoK}
\frac{r}{R}= 1+ W_0\left(-\frac{X^+X^-}{4R^2 }\right)\ ,
\eeq
with $W_0$ the Lambert W function, showing that the boundary of the Rindler region is time-dependent in the local Minkowski coordinates.  

The equation of motion may also be studied in these coordinates, and for a mode with definite angular momentum 
\beq
\phi_{lm} = u_l \frac{Y_{lm}(\Omega)}{r^{D/2-1}}
\eeq
becomes
\beq\label{EOMKruskal}
\partial_{X^+}\partial_{X^-}u_l = -\frac{1}{4} \frac{R}{r} e^{-r/R+1} V_l(r)u_l
\eeq
with $V_l(r)$ given by \eqref{4dpot} together with \eqref{rtoK}.  Notice that as a result of the latter equation, the effective potential is time-dependent in the locally Minkowski coordinates.  While the general form of solutions appears less transparent in these coordinates, the solutions do simplify when restricted to the Rindler region.  In this region, \eqref{EOMKruskal} becomes the 2d massive wave equation,
\beq
4\partial_{X^+}\partial_{X^-}u_l = -m_l^2 u_l\ ,
\eeq
with effective mass term
\beq
m_l^2=\frac{l(l+1)+1}{ R^2}\ .
\eeq
This has a basis of solutions (see \eqref{KruskalModes})
\beq\label{Ksolns}
u_{kl}= e^{ikX-i\omega_k T}\quad , \quad \tilde u_{kl}= e^{-ikX-i\omega_k T}
\eeq
with $\omega_k^2=k^2+m_l^2$\ .
These are neither purely outgoing or ingoing, but do become purely outgoing or ingoing, respectively, in the large $k$ limit.  
One may also compare the equations and solutions in the Eddington--Finkelstein coordinates; from \eqref{Kcoorddef}, 
the equation \eqref{EOMKruskal} becomes
\beq
\partial_{x^+}\partial_{x^-}u_l = -\frac{1}{4}\left(1-\frac{R}{r}\right) V_l(r) u_l\ ,
\eeq
and likewise inside the horizon, in terms of $\hat x^-$.  In the Rindler region, the effective potential in these coordinates vanishes, resulting in solutions of the form $e^{-i\omega x^\pm}$.

Comparing these descriptions provides another way to compute the Bogolubov coefficients, in this high momentum, near horizon limit. The approximate solutions \eqref{Ksolns} are related to the energy eigenmodes \eqref{EEigenmodes} by
\beq
u_{\omega}= \int dk \big( \alpha^+_{\omega k} u_{k}+\alpha^-_{\omega k}u_{k}^*\big) ,
\eeq
where the spherical indices have been suppressed to simplify the notation. The Bogolubov coefficients can be calculated by performing the Fourier transform in the usual way
\beq
\alpha^+_{\omega k}=\frac{1}{4\pi k} (u_k,u_\omega) = \frac{i}{4\pi k} \int dX^- \big(u_{k}^*\partial_{X^-}u_\omega - \partial_{X^-}u_{k}^* u_\omega \big),
\eeq
where it is useful to take the inner product on a null surface with coordinate $X^-$. Similarly, the other coefficient is $\alpha^-_{\omega k}=-(u_k^*,u_\omega)/(4\pi k)$. The resulting integrals are
\bea
\alpha^+_{\omega k}&=&\frac{1}{2\pi i k} \left(\frac{1}{2ikR}\right)^{2i\omega R}\Gamma(1+2 i \omega R)\ ,\cr
\alpha^-_{\omega k}&=&-\frac{1}{2\pi i k}\left(\frac{1}{-2ikR}\right)^{2i\omega R}\Gamma(1+2 i \omega R)\ .
\eea 
This is of the same form as  Hawking's result from \cite{Hawk}, modulo conventions.

\mciteSetMidEndSepPunct{}{\ifmciteBstWouldAddEndPunct.\else\fi}{\relax}
\bibliographystyle{utphys}
\bibliography{dynBH}{}

\end{document}